\documentclass[11pt]{article}
\usepackage{times}
\usepackage{geometry}
\usepackage[utf8]{inputenc}
\usepackage{enumitem,amssymb}
\usepackage{ragged2e}
\newlist{thematic}{itemize}{8}
\setlist[thematic]{label=$\square$}
\usepackage{pifont}
\usepackage{fancyhdr}
\usepackage[USenglish]{babel}
\usepackage[utf8]{inputenc}
\usepackage[T1]{fontenc}
\usepackage{epsfig}
\usepackage{wrapfig}
\usepackage{color}

\usepackage[vflt]{floatflt}
\usepackage{longtable}
\usepackage{caption}
\usepackage{jheppub}   
\usepackage[dvipsnames,svgnames,x11names]{xcolor}
\usepackage[all]{hypcap} 
\usepackage{amssymb,amsmath}
\usepackage{longtable}
\usepackage{amssymb}
\usepackage{amsmath, xspace}
\usepackage{graphics,graphicx} 
\usepackage{rotating}
\usepackage{color}
\usepackage{xcolor}
\usepackage{multirow}
\usepackage{subfigure}
\usepackage{sectsty}
\usepackage[outercaption]{sidecap}
\sidecaptionvpos{figure}{c}
\usepackage{wrapfig}

\definecolor{DarkGreen}{rgb}{0.0, 0.3, 0.0}
\definecolor{purple}{rgb}{0.5, 0.0, 0.5}
\definecolor{red}{rgb}{1, 0.0, 0.0}
\definecolor{green}{rgb}{0, 1.0, 0.0}

\def\3he{$^3{\rm He}$}

\def\lsim{\mathrel{\lower2.5pt\vbox{\lineskip=0pt\baselineskip=0pt
           \hbox{$<$}\hbox{$\sim$}}}}

\def\gsim{\mathrel{\lower2.5pt\vbox{\lineskip=0pt\baselineskip=0pt
           \hbox{$>$}\hbox{$\sim$}}}}

\begin{document}
\huge
\begin{center}
AtLAST - Cosmology with submillimetre galaxies\\
magnification bias\\
{\normalsize This white paper was submitted to ESO Expanding Horizons in support of AtLAST}
\end{center}

\bigskip
\normalsize

\textbf{Authors:} 
Laura Bonavera$^{1,2}$ (bonaveralaura@uniovi.es); Joaquin Gonzalez-Nuevo$^{1,2}$; Juan Alberto Cano$^{1,2}$; David Crespo$^{1,2}$; Rebeca Fernández-Fernández$^{1,2}$; Valentina Franco$^{1,2}$; Marcos M. Cueli$^{1,2}$; José Manuel Casas$^{1,2}$; Tony Mroczkowski$^3$; Caludia Ciccone$^4$; Marina Migliaccio$^{5,6}$; Evanthia Hatziminaoglou$^{7,8,9}$, Hugo Messias$^{10,11}$\\
$^1$Departamento de Fisica, Universidad de Oviedo, C. Federico Garcia Lorca 18, 33007 Oviedo, Spain.\\
$^2$Instituto Universitario de Ciencias y Tecnologías Espaciales de Asturias (ICTEA), C. Independencia 13, 33004 Oviedo, Spain.\\
$^3$Institute of Space Sciences (ICE, CSIC), Carrer de Can Magrans, s/n, 08193 Cerdanyola del Vallès, Barcelona, Spain.\\
$^4$Institute of Theoretical Astrophysics, University of Oslo, Postboks 1029, Blindern, 0315 Oslo\\
$^5$Dipartimento di Fisica, Università di Roma Tor Vergata, Via della Ricerca Scientifica 1, 00133 Roma, Italy\\
$^6$INFN Sezione di Roma2, Università di Roma Tor Vergata, Via della Ricerca Scientifica 1, 00133 Roma, Italy\\
$^7$ESO, Karl-Schwarzschild-Str. 2, 85748 Garching bei München, Germany\\
$^8$Instituto de Astrof\'isica de Canarias (IAC), E-38200 La Laguna, Tenerife, Spain\\
$^9$Departamento de Astrof\'isica, Universidad de La Laguna, E-38206 La Laguna, Tenerife, Spain\\
$^{10}$European Southern Observatory, Alonso de C\'ordova 3107, Vitacura, Casilla 19001, Santiago de Chile, Chile\\
$^{11}$Joint ALMA Observatory, Alonso de C\'ordova 3107, Vitacura 763-0355, Santiago, Chile\\


\textbf{Science Keywords:} 
cosmology: large-scale structure of universe, cosmology: dark energy, cosmology: dark matter, cosmology: lenses, cosmology: observations

 \captionsetup{labelformat=empty}
\begin{figure}[h]
   \centering
\includegraphics[width=.49\textwidth]{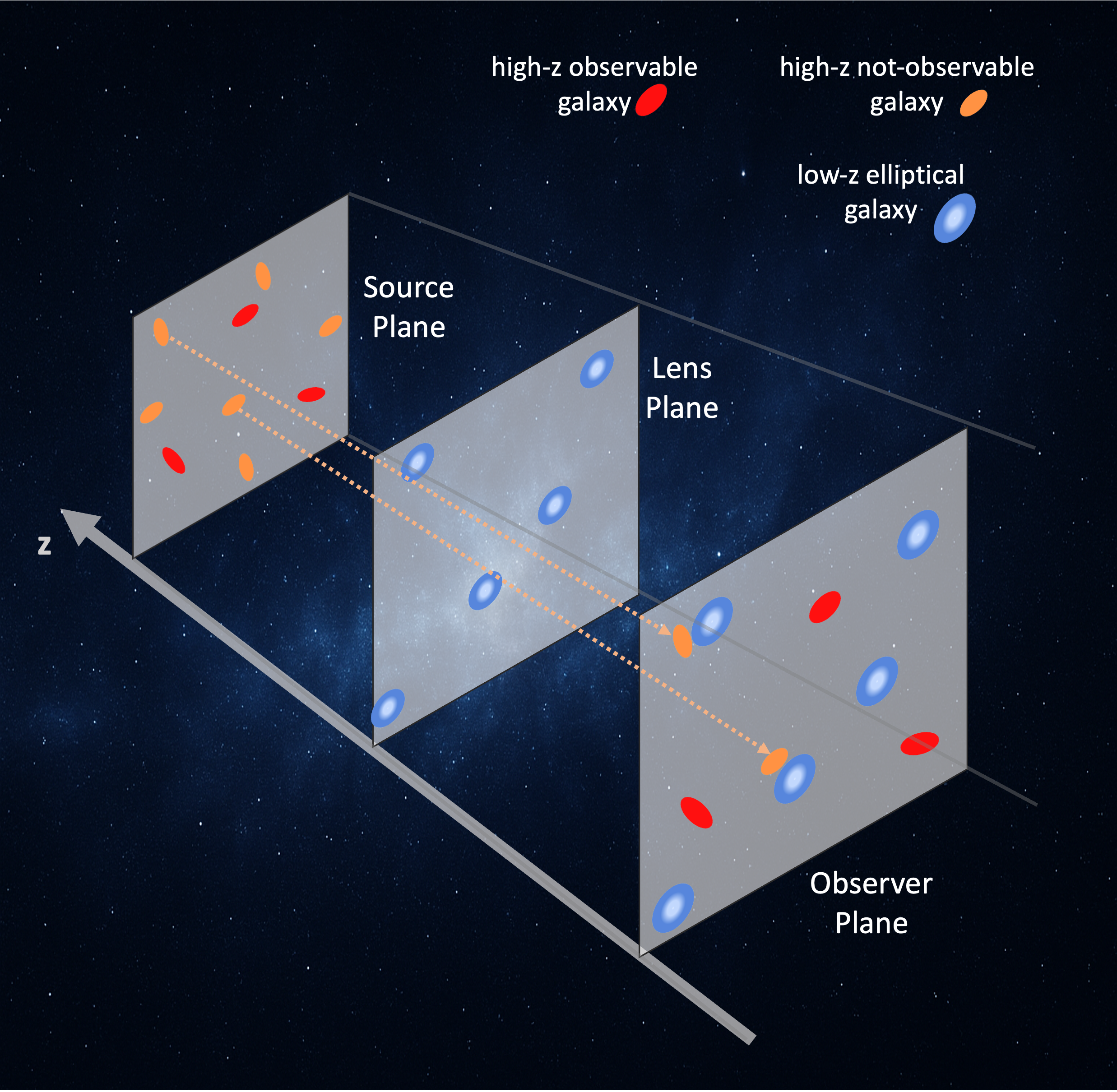}
   \caption{Description of the Magnification Bias Effect}
\end{figure}
\vspace{-15mm}

\setcounter{figure}{0}
\captionsetup{labelformat=default}


\pagebreak
\section*{Abstract}

Magnification bias offers a uniquely powerful and independent route to cosmological information. As a gravitational-lensing observable, it probes galaxy–matter correlations without relying on galaxy shapes, PSF modelling, or intrinsic-alignment corrections. Its sensitivity spans both geometry and growth: magnification bias simultaneously responds to the matter density, the amplitude of structure, and the redshift evolution of dark energy (DE) below $z\leq 1$. Importantly, its parameter degeneracy directions differ from those of shear, Baryon Acoustic Oscillations (BAO), and Cosmic Microwave Background (CMB) data, making it a complementary and consistency-check probe with substantial diagnostic value for the next decade of precision cosmology. However, the current potential of magnification bias is restricted by limited sky coverage, catalogue inhomogeneities, and insufficiently precise redshift or number-count characterisation. A next-generation wide-field submillimetre facility like AtLAST—capable of uniform, deep surveys and spectroscopic mapping—would overcome these limitations and transform magnification bias into a competitive, high-precision cosmological tool.

\section{Scientific context and motivation}

Weak gravitational lensing alters the apparent properties of distant galaxies through both shape distortions (shear) and changes in their observed flux and angular size (magnification, Schneider et al. 1992). Magnification bias refers to the resulting mismatch between intrinsic and observed number counts of background sources: flux boosting brings intrinsically faint sources above the detection threshold, while solid-angle dilation reduces their surface density. The balance between these competing effects is governed by the logarithmic slope of the source number counts, $\beta$. When $\beta>1$—as is the case for high-redshift submillimetre galaxies (SMGs)—flux boosting dominates and produces an excess of background sources around foreground mass overdensities. The statistical measurement of this effect through foreground–background cross-correlations is known as cosmic magnification (Bartelmann \& Schneider 2001).

Cosmic magnification is a gravitational lensing observable that is independent and complementary to the more widely exploited shear-based techniques. It probes the galaxy–matter cross-correlation without requiring resolved galaxy shapes, PSF modelling, intrinsic-alignment mitigation, or shear calibration. This simplicity makes magnification bias particularly powerful for dusty, high-redshift galaxies, for which shape measurements are challenging or impossible. It also decorrelates the degeneracy directions characteristic of shear observables: magnification bias is simultaneously sensitive to the matter density $\Omega_m$ and the amplitude of matter fluctuations $\sigma_8$, and it shows strong sensitivity to DE evolution at $z\leq1$ ( Bonavera et al. 2021, Cueli et al. 2024a). As a result, magnification bias provides cosmological information otherwise inaccessible, while also serving as a robustness cross-check for analyses dominated by optical shear surveys such as \href{https://www.darkenergysurvey.org}{\underline{DES}} (Troxel et al., 2018), \href{https://kids.strw.leidenuniv.nl}{\underline{ KiDS}} (Hildebrandt et al., 2017), and \textit{Euclid} (Schrabback, T. et al. 2025).

Over the past decade, the availability of large samples of \textit{Herschel}-selected SMGs has enabled the first cosmological studies using magnification bias. Cross-correlations between low-redshift lenses (e.g. Galaxy and Mass Assembly, GAMA; Driver et al. 2011) and high-redshift SMGs from \textit{Herschel} Astrophysical Terahertz Large Area Survey (H-ATLAS; Eales et al. 2010; Smith et al. 2017) have yielded constraints on $\Omega_m$, $\sigma_8$, and, through tomographic analyses, the DE equation-of-state parameters $w_0$ and $w_a$ ( Bonavera et al. 2021, Cueli et al. 2024a). These results have provided hints of DE evolution consistent with recent \href{https://www.desi.lbl.gov}{\underline{DESI}} BAO findings, have produced upper limits on the sum of neutrino masses, and have allowed detailed modelling of foreground halo masses and occupation statistics. Importantly, magnification constraints on $\Omega_m$ and $\sigma_8$ follow different degeneracy directions from shear, offering valuable complementarity for joint analyses.

Yet, despite this promise, magnification bias remains statistically limited. Present constraints are dominated by restricted sky coverage ($\sim200$ deg$^2$ for the GAMA/H-ATLAS overlap), catalogue inhomogeneities, and sampling-variance biases that can shift cosmological posteriors. The need to fix $\beta$ with a strong prior reflects the limited depth of current continuum data, while the small angular scales accessible in \textit{Herschel} fields suppress sensitivity to large-scale modes that carry much of the cosmological information. Together, these limitations have constrained magnification bias to a niche role, despite its conceptual simplicity and its unique cosmological leverage. A next-generation transformative submillimetre (submm) facility is thus required to unlock the true potential of magnification bias (Fernandez-Fernandez et al., 2025). 

\vspace{-3mm}
\section{Science case}
Magnification bias, statistically quantified as cosmic magnification, has emerged as an independent and complementary tool for cosmology, particularly since the availability of large samples of high-redshift SMGs from the \textit{Herschel} observatory. The key results derived from magnification bias studies include constraints on the standard cosmological parameters, limitations on the sum of neutrino masses, and evidence regarding DE evolution.

\begin{wrapfigure}{r}{0.4\textwidth}
\includegraphics[width=.9\linewidth]{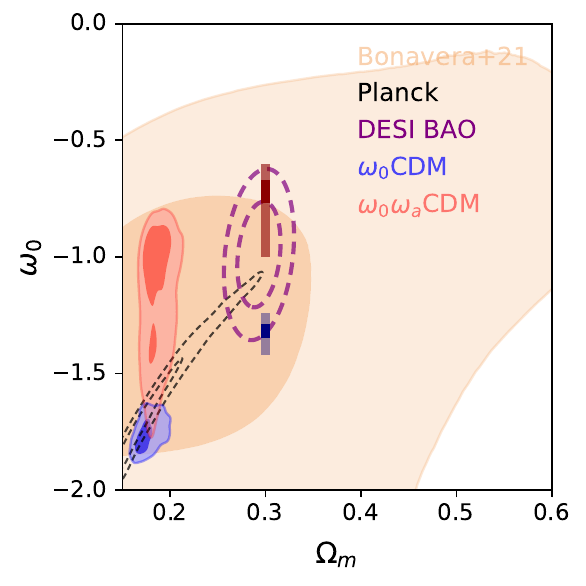}
   \caption{Preliminary $w_0$-$\Omega_m$ posterior probability contours for different study cases with and without fixing $\Omega_m$. Data from other works are also plotted, namely Bonavera et al. (2021), Planck Collaboration (2020) and approximate Gaussian contours from DESI BAO in the $\omega_0$CDM model (DESI Collaboration, 2025a).}
\end{wrapfigure}

\textbf{Improving cosmological constraints within $\Lambda$CDM with an independent cosmological probe.} Magnification bias is sensitive to both the matter density parameter ($\Omega_m$) and the amplitude of the smoothed matter overdensity field ($\sigma_8$). Latest results produce independent competitive constraints on both quantities ($\Omega_m=0.27^{+0.02}_{-0.04}$ and $\sigma_8=0.72\pm0.04$), without the typical degeneracy between them (Cueli et al., 2024a). A wide-area, homogeneous, high-resolution continuum survey would significantly reduce sampling variance, measure $\beta$ directly, and allow background tomographic analysis, leading to improved constraints on $\sigma_8$, $\Omega_m$ both bellow <5\%, and opening the possibility of constraining also $H_0$.

\textbf{Is DE evolving?} The latest DESI results (DESI Collaboration, 2025b) show a strong preference, $4.2\sigma$ at the highest, for a dynamical DE model over the static cosmological constant ($\Lambda$CDM). All combinations of available datasets favour results where the parameters lie in the quadrant with $w_0>-1$ and $w_a<0$. Magnification bias is particularly sensitive to DE evolution at $z\leq1$, as demonstrated by its latest results ($w_0=-1.09^{+0.75}_{-0.63}$ and $w_a=-0.19^{+1.67}_{-1.69}$; Bonavera et al.,in prep), in agreement with DESI scenario. 
Assuming an order-of-magnitude increase in sky area and a factor 5–10 increase in background source counts, magnification bias constraints can achieve $\sigma(w_0) \sim 0.1$ and meaningful $w_a$ detection capability, with $\sigma(w_a) \sim 0.3-0.5$.

\textbf{Tensions on Neutrino Mass ($\Sigma m_\nu$).} DESI data provides tight constraints on the sum of neutrino masses when combined with CMB results yielding the tightest upper bound to date, $\Sigma m_\nu<0.064$ eV (95\% limit; DESI Collaboration, 2025b). This upper limit is very close to the lower bound set by neutrino oscillation experiments $\Sigma m_\nu\geq0.059$ eV and it also shows a posterior mass distribution peaking at non-physical negative ‘effective’ neutrino masses if allowed. Magnification bias is one of the few observables capable of constraining the total neutrino mass on it owns. Using a single redshift wide-bin and assuming a primordial power spectrum amplitude it yields $\Sigma m_\nu<0.78$ eV (95\% limit; Cueli et al., 2024b). Considering an order-of-magnitude increase in sky area and a factor 5–10 increase in background source counts, magnification bias through a tomography analysis for both foreground and background sample could produce tighter constraints by a factor 5-10.

To establish magnification bias as a precision cosmological probe, we need a facility that combines two key capabilities: extremely low confusion noise, which requires a large aperture, and the ability to conduct wide-area, homogeneous surveys at high speed, which demands a large field of view. These features must operate in the submm regime to detect and characterize the full, unbiased population of highly dust-obscured sources. Current instruments fall short of this combination: \href{https://almascience.eso.org}{\underline{ALMA}} excels at small-field, high-resolution imaging, while facilities such as \href{https://www.eaobservatory.org/jcmt/instrumentation/continuum/scuba-2/}{\underline{SCUBA-2}}, \href{https://pole.uchicago.edu/public/Home.html}{\underline{SPT-3G}}, \href{https://simonsobservatory.org}{\underline{SO}}, and planned 6-m class telescopes (e.g. \href{https://www.ccatobservatory.org}{\underline{FYST}}) prioritize wide-field mapping at low resolution. Optical and near-infrared observatories like \href{https://www.cosmos.esa.int/web/euclid}{\textit{\underline{Euclid}}} and \href{https://www.lsst.org}{\underline{LSST}}, meanwhile, are largely blind to the source population critical for this cosmological probe.

The Atacama Large Aperture Submillimeter Telescope (\href{https://www.atlast.uio.no}{\underline{AtLAST}}; Mroczkowski et al., 2025) will overcome key limitations in magnification bias studies, especially those related to observational scope, sample size, homogeneity, and statistical precision. With wide, homogeneous sky coverage, unprecedented angular resolution, and deep spectroscopic capability in the submm band, it will remove the statistical and systematic limitations that have constrained this technique for decades. The resulting data will provide independent constraints on DE evolution, matter density, and structure growth, while deepening our understanding of galaxy–halo connections and the high-redshift universe. Thus, AtLAST will transform magnification bias from a secondary probe into a cornerstone of precision cosmology in the 2030s and beyond.

\section{Technical requirements}
AtLAST exemplifies the type of facility required to advance magnification bias as a precision cosmological probe (van Kampen et al., 2025). It is a proposed 50-meter single-dish telescope located at a high-altitude site near ALMA ($\sim 5000$ m; Reichert et al., 2024), operating across the full submm range from 30 to 950 GHz (0.3-10 mm; Mroczkowski et al., 2025). Its large aperture delivers exceptional sensitivity and diffraction-limited resolution of about $1.5''$ at 950 GHz, effectively eliminating confusion noise and enabling the detection of faint, dust-obscured galaxies. Complementing this, AtLAST offers an instantaneous field of view of $2^\circ$, which is critical for high-speed, wide-area surveys and for recovering extended low-surface-brightness structures that interferometers cannot capture (Gallardo et al., 2024).

The combination of a large collecting area and wide field of view results in mapping speeds more than three orders of magnitude faster than ALMA, while maintaining continuum sensitivity comparable to the entire ALMA array post-upgrade. AtLAST will host next-generation instruments with up to $10^6$ detector elements, enabling simultaneous spectroscopic and continuum observations (Gallardo et al., 2024). Uniquely, it pioneers sustainable astronomy through a fully off-grid renewable energy system with hybrid battery and hydrogen storage, ensuring 100\% power availability for day and night operations (Viole et al., 2023; Kiselev et al., 2024). In short, AtLAST is not just a telescope, it is a transformative platform that will redefine submm cosmology and deliver science that no other planned facility can achieve.

\bigskip
\noindent\textbf{References:} \small{Bartelmann, M. \& Schneider, P. 2001, Phys. Rep., 340, 291  $\bullet$ Bonavera, L., et al. 2021, A\&A, 656, A99 $\bullet$ Cueli, M.M., et al. 2024a, A\&A, 686, A190 $\bullet$ Cueli M.M., et al., 2024b, A\&A, 687, A300 $\bullet$ DESI Collaboration et al., 2025a, JCAP, 02, 021 $\bullet$ DESI Collaboration et al., 2025b, Phys. Rev. D, 112, 083515 $\bullet$ Driver, S. P., et al. 2011, MNRAS, 413, 971 $\bullet$ Eales, S., et al. 2010, PASP, 122, 499 $\bullet$ Fernandez-Fernandez, R., et al., 2025, arXiv:2510.23582 $\bullet$ Gallardo, P.A., et al., 2024, Proc. of the SPIE, 13094, 1309428 $\bullet$ Hildebrandt et al., 2017, MNRAS, 465, 1454 $\bullet$ van Kampen, E., et al., 2025, Open Research Europe, 4, 122 $\bullet$ Kiselev, A., et al., 2024, Proc. of the SPIE, 13094, 130940E $\bullet$ Mroczkowski, T., et al., 2025, A\&A, 694, A142 $\bullet$ Nayyeri, H., et al., 2016, ApJ, 823, 17 $\bullet$ Planck Collaboration, et al., 2020, A\&A, 641, A6 $\bullet$ Reichert, M., et al., 2024, Proc. of the SPIE, 13094, 130941U $\bullet$ Schneider, P., Ehlers, J., \& Falco, E. E. 1992, Gravitational Lenses $\bullet$ Schrabback, T., et al. 2025, arXiv:2507.07629 $\bullet$ Smith, M. W. L., et al. 2017, ApJS, 233, 26 $\bullet$ Troxel et al., DES Collaboration, 2018, Phys. Rev. D, 98, 043528 $\bullet$ Viole, I., et al., 2023, Energy, 282, 128570}

\end{document}